\documentclass[preprint,aps,prd,showpacs,nofootinbib,superscriptaddress,tightenlines,floatfix]{revtex4-1}

\usepackage{amsmath}
\usepackage{graphicx}
\usepackage{subfigure}
\usepackage{epstopdf}
\usepackage{epsfig}
\usepackage{amssymb}
\usepackage{bm}
\usepackage{bbm}
\newcommand{\beq}{\begin{equation}}
\newcommand{\eeq}{\end{equation}}

\newcommand{\pa}{\partial}

\usepackage{color}

\begin{document}

\begin{flushright}
{\small KIAS-PREPRINT-P14030} 
\end{flushright}

\title{Dark Matter and Dark Force in the Type-I Inert 2HDM \\
 with Local $U(1)_H$ Gauge Symmetry
}

\author{P. Ko}
\affiliation{School of Physics, KIAS, Seoul 130-722, Korea}

\author{Yuji Omura}
\affiliation{Department of Physics, Nagoya University, Nagoya 464-8602, Japan}

\author{Chaehyun Yu}
\affiliation{School of Physics, KIAS, Seoul 130-722, Korea}


\begin{abstract} 
We discuss dark matter (DM) physics in the Type-I inert two-Higgs-doublet model (2HDM) 
with local $U(1)_H$ Higgs gauge symmetry, which is  
assigned to the extra Higgs doublet in order to avoid the Higgs-mediated flavor problems.  
In this gauged inert DM setup, a $U(1)_H$-charged scalar $\Phi$ is also introduced to break $U(1)_H$ 
spontaneously through its nonzero vacuum expectation value (VEV), $\langle \Phi \rangle$, and then
the remnant discrete subgroup appears according to the $U(1)_H$ charge assignment of $\Phi$.
The $U(1)_H$-charged Higgs doublet does not have Yukawa couplings with the 
Standard-Model (SM) fermions, and its lightest neutral scalar component $H$ is stable 
because of the remnant discrete symmetry.
In order to suppress a too large $Z$-exchange diagram contribution 
in DM direct detection experiments, 
we have to introduce a non-renormalizable operator which can be generated by integrating out an extra heavy scalar.
With these new particles contents, we first investigate the constraint on the 
$U(1)_H$ gauge interaction, especially through the kinetic and mass mixing  between 
the SM gauge bosons and the extra gauge boson. 
Then we discuss dark matter physics in our 2HDM: thermal relic density, and direct/indirect 
detections of dark matter.  The additional $U(1)_H$ gauge interaction plays a crucial role  
in reducing the DM thermal relic density.  The most important result within the inert DM 
model with local $U(1)_H$ symmetry is that $\sim O(10)$ GeV dark matter scenario, 
which is strongly disfavored in the usual Inert Doublet Model (IDM) with
$Z_2$ symmetry, is revived in our model because of newly open channels,  
$H H \rightarrow Z_H Z_H , Z_H Z$. 
Exotic Higgs decays, $h\to Z_H Z_H, Z Z_H$, would be distinctive signatures of 
the inert 2HDM with local $U(1)_H$ symmetry.
\end{abstract}


\maketitle

\section{Introduction}
The discovery of a Standard-Model-like Higgs boson at the LHC ~\cite{Aad:2012tfa,Chatrchyan:2012ufa} opens new era in particle physics and cosmology. 
The precise measurements of its mass and couplings to the Standard Model (SM)
particles will reveal the structure of the Higgs sector, which is the least known piece 
in the SM.  Up to now, its couplings to the ordinary particles are consistent with
the predictions of the SM within uncertainties and most of results at the LHC can be 
understood in the framework of the SM~\cite{Aad:2013xqa,Chatrchyan:2013mxa}.
On the other hand, there are some clues on new physics beyond the SM:
nonbaryonic dark matter (DM), dark energy, neutrino oscillation, baryon asymmetry 
of the universe, and etc., which cannot be explained by the renormalizable SM 
and require its extensions beyond the SM. 

One simple extension of the SM is to add one extra Higgs doublet.
In fact, many high-energy theories predict extra Higgs doublets, 
and the two-Higgs-doublet models (2HDMs) could be interpreted as the effective theories 
of those high energy theories after we integrate out heavy particles.
Of course, the 2HDMs could be interesting by themselves,  because of their rich 
phenomenology and benchmark models with an extended Higgs sector. 
2HDMs predict extra neutral and charged scalar bosons in addition to a SM-like Higgs 
boson, and the extra scalar bosons may change phenomenology of Higgs boson and 
SM particles at colliders~\cite{rich}.
We could also find cold dark matter candidates in some 2HDMs:
one of the extra scalars~\cite{IDM} or one of the extra fermions, which
may be added to the models in 2HDMs with gauged $U(1)_H$ symmetry 
~\cite{Ko-2HDM}.   When the Higgs potential in 2HDMs has 
a CP-violating source, the baryon asymmetry of the universe may be 
explained~\cite{BAU}.  
And small neutrino masses may naturally be generated by one-loop diagram
in some 2HDMs~\cite{neutrino}.
Finally, it is very interesting that in 2HDMs with flavor-dependent  
$U(1)_H$ gauge symmetry, the anomalies in the top forward-backward
asymmetry at the Tevatron and $B\to D^{(\ast)}\tau\nu$ decays
at BABAR may be reconciled~\cite{chiralU1}.

One important phenomenological issue in models with extra Higgs doublets is 
the so-called Higgs--mediated flavor changing neutral current (FCNC) problem.
If a right-handed (RH) fermion couples with more than two Higgs doublets,
FCNCs involving the neutral scalar bosons  
generally appear after the electro-weak (EW) symmetry breaking.
In many cases, this Higgs--mediated flavor problem is resolved by imposing  
softly-broken $Z_2$ symmetry  \'{a} la Glashow and Weinberg~\cite{Glashow}. 
The discrete $Z_2$ symmetry could be replaced by other discrete 
symmetry~\cite{s3} or continuous local gauge symmetry~\cite{Ko-2HDM}.  
In fact, the present authors proposed a new class of  2HDMs where $U(1)_H$ Higgs 
gauge symmetry is introduced  instead of  softly broken discrete $Z_2$ symmetry, 
in order to avoid the flavor problem \cite{Ko-2HDM}, and discussed the phenomenology 
of Type-I 2HDM with $U(1)_H$ in Ref. \cite{Ko-2HDM1}. When Higgs doublets are charged 
under the extra gauge symmetry, the so-called $\rho$ parameter is deviated from the
SM prediction at the tree level if the $U(1)_H$-charged Higgs doublet develops 
a nonzero vacuum expectation value (VEV).
In Ref. \cite{Ko-2HDM1}, the authors investigate the constraints  not only from 
the EW precision observables (EWPOs) but also from the recent LHC results 
on the SM-like Higgs search,  especially in Type-I 2HDMs where only one Higgs 
doublet is charged and the SM particles are neutral under $U(1)_H$. 
The deviation of the $\rho$ parameter is mainly  from the tree-level mass mixing 
between $Z$ and the $U(1)_H$ gauge boson ($Z_H$),  and from the mass differences 
among the scalar bosons. The bounds  on EWPOs require small $U(1)_H$ interactions, 
so that its effects become tiny in physical observables.

In this work, we consider a new scenario where the VEV of the $U(1)_H$-charged Higgs doublet is zero 
in our Type-I 2HDM and a $U(1)_H$-charged SM-singlet scalar $\Phi$  breaks $U(1)_H$ spontaneously.
In this case, the mass mixing between gauge bosons is also negligible at the tree level.   
Therefore we can expect that  the $Z_H$ gauge interaction becomes sizable 
and then  the idea of gauged $U(1)_H$ Higgs symmetry might be tested.  
The bounds on $Z_H$ mass and its couplings to the SM particles will come from the loop-level mass 
and kinetic mixings between the SM gauge bosons and $Z_H$, as well as the tree-level kinetic mixing.

On the other hand, the SM fermions are chiral under new $U(1)_H$ gauge 
symmetry, so that the model may be anomalous unless new chiral fermions are introduced. 
As discussed in Ref. \cite{Ko-2HDM,Ko-2HDM1}, we could consider the anomaly-free $U(1)_H$
charge assignment to build the 2HDM with local Higgs symmetry: For example, the SM particles are
not charged and only one extra Higgs doublet is charged under $U(1)_H$.
In our 2HDM, the $U(1)_H$ will be broken by   the nonzero VEV of $\Phi$.
However, the residual symmetry may still remain and we could find 
stable cold dark matter (CDM) candidates. 
In Sec. \ref{sec3}, we introduce the conditions for the stability of DM candidates and 
discuss dark matter physics in the Type-I 2HDMs with local $U(1)_H$ gauge symmetry.
One well-known dark matter model among 2HDMs is the so-called inert doublet model 
(IDM)~\cite{IDM,Ma}.   In the ordinary IDM, one extra $Z_2$ symmetry is imposed and the extra 
Higgs doublet is the only $Z_2$-odd particle. 
If the Higgs doublet does not develop the nonzero VEV, $Z_2$ symmetry forbids the decay of the 
lightest $Z_2$-odd scalar boson,  and the scalar boson could be a good CDM candidate. 
The scalar CDM interacts with the SM particle through the scalar exchange and the EW 
interaction, and one could find the favored regions for the correct thermal relic density 
around $m_{\rm DM} \sim 60$ GeV or $m_{\rm DM} > 500$ GeV \cite{IDM},  
which are safe for the constraints from the collider searches and the DM direct detection searches. 
Especially, the light CDM scenario faces the strong bound from the invisible search 
at the LHC,  and the only allowed mass region is around the resonance of the SM-Higgs.

In our 2HDM with local $U(1)_H$ gauge symmetry, at least one 
Higgs doublet  is charged under the $U(1)_H$ symmetry, and  the scalar component 
respects the remnant discrete symmetry  of $U(1)_H$ after the EW and $U(1)_H$ 
symmetry breaking.  This discrete symmetry originates from the local $U(1)_H$ and will
protect the DM from decay to all orders in perturbation theory even in the presence of 
higher dimensional operators. 
Stability of the scalar DM is guaranteed by  the local discrete symmetry, and the scalar 
DM interacts with the SM particles strongly through the $U(1)_H$ gauge interaction.
The extension of the usual IDM to the $U(1)_H$ gauge symmetric one not only suggests
the origin of the $Z_2$ Higgs symmetry, but also may open up a new scenario for 
dark matter phenomenology,  which cannot be achieved in  the usual IDM. 
In fact, the  $U(1)_H$ gauge interaction can plays an important role in thermalizing  
the scalar DM,  and  $\sim O(10)$ GeV CDM scenario could be revived in our IDM, 
which is  a very interesting aspect of our model.
In Sec. \ref{sec3-3} and Sec. \ref{sec3-4}, we investigate the constraints on 
not only the relic density and the DM direct detection, but also the DM indirect detection, 
in the IDM with $Z_2$ symmetry (IDMw$Z_2$) and  
in the IDM with local $U(1)_H$ symmetry (IDMw$U(1)_H$).
The indirect astrophysical observations would be one of the ways to prove our model  
so that we calculate the velocity-averaged cross section for dark matter
annihilation 
in the halo, 
and consider the constraints from the Fermi-LAT
by observing the $\gamma$-ray flux from the dwarf spheroidal galaxies, 
which is one of the recent results relevant to the light CDM scenario.

The organization of this paper is as follows.
In Sec. \ref{sec2}, we introduce the Lagrangian of the Type-I 2HDM with $U(1)_H$ Higgs symmetry
and discuss the constraints on the $U(1)_H$ gauge interactions.
In Sec. \ref{sec3}, we discuss dark matter physics in the IDMw$Z_2$ and IDMw$U(1)_H$:
the stability of CDM, the thermal relic density, the DM direct and indirect detections.
Sec. \ref{sec4} is devoted to our conclusion.

\section{2HDM with $U(1)_H$ Higgs symmetry}
\label{sec2}
Based on Ref. \cite{Ko-2HDM1}, we discuss our setup of the Type-I 2HDM with $U(1)_H$ 
Higgs symmetry, where only extra Higgs doublet $H_1$ is charged under the $U(1)_H$ 
gauge symmetry and all the SM particles including the SM Higgs doublet
$H_2$ are neutral.  
The Lagrangian for gauge fields and scalar fields is given by 
\begin{eqnarray}
{\cal L}  &=&  -\frac{1}{4} F^{\mu \nu}_Y F_{Y\mu \nu}  -\frac{1}{4} F^{a\mu \nu}_{\rm SU(2)_L} F^a_{{\rm SU(2)_L} \mu \nu} -\frac{1}{4} F^{\mu \nu}_H F_{H\mu \nu} -\frac{\kappa}{2} F_Y^{\mu \nu} F_{H\mu \nu}  \nonumber \\
&& + \left | (\pa_{\mu} -\frac{i}{2}g'B_{\mu}  -\frac{i}{2}gA^a_{\mu} \tau^a-iq_{H_1}g_H \Hat{Z}_{H\mu})H_1 \right|^2 \nonumber \\ 
&&+ \left|( \pa_{\mu} -\frac{i}{2}g'B_{\mu}  -\frac{i}{2}gA^a_{\mu} \tau^a)H_2 \right|^2
+\left | (\pa_{\mu} -iq_{\Phi}g_H \Hat{Z}_{H\mu})\Phi \right|^2 
- V_{\rm scalar}(H_1,H_2,\Phi), 
\label{eq:lagrangian}%
\end{eqnarray}
where $F^{\mu\nu}_Y$, $ F^{a\mu \nu}_{\rm SU(2)_L}$, and $ F^{\mu \nu}_H$ are the 
field strengths of $U(1)_Y$, $SU(2)_L$, and $U(1)_H$ of the gauge fields, $B^{\mu}$, $A^a_{\mu}$,
and $\Hat{Z}^{\mu}_H$, and  $g'$, $g$ and $g_H$ are their gauge couplings. 
$q_{H_1}$ is the $U(1)_H$ charge of $H_1$. 
$\kappa$ is the kinetic mixing between $U(1)_Y$ and $U(1)_H$ field strength tensors, 
which is allowed by the local gauge symmetry. It is assumed to be a free parameter. 
Finally $V_{\rm scalar}(H)$ is the potential for the complex scalars: 
\begin{eqnarray}
V_{\rm scalar}&=& (m_1^2+ \widetilde{\lambda}_1 |\Phi|^2) H^{\dagger}_1 H_1+ (m_2^2+ \widetilde{\lambda}_2 |\Phi|^2) H^{\dagger}_2 H_2    \nonumber \\
&&+ \frac{\lambda_1}{2} (H^{\dagger}_1 H_1)^2 + \frac{\lambda_2}{2} (H^{\dagger}_2 H_2)^2 
+ \lambda_{3} (H^{\dagger}_1 H_1)(H^{\dagger}_2 H_2)+ \lambda_{4} |H^{\dagger}_1 H_2|^2 \nonumber \\
&& +m^2_{\Phi} |\Phi|^2 + \lambda_{\Phi} |\Phi|^4+ c_l \left ( \frac{\Phi}{\Lambda_\Phi} \right )^l (H^{\dagger}_1 H_2)^2+h.c..
\label{eq:potential}
\end{eqnarray}
A new scalar  $\Phi$ is a singlet under the SM gauge group, but  is charged under 
$U(1)_H$ gauge symmetry,  and thus it breaks the $U(1)_H$ by the nonzero VEV 
$(\langle \Phi \rangle \equiv v_{\Phi})$.  $c_l$ is the dimensionless coupling of the 
higher-dimension operator, which is suppressed by one arbitrary scale $(\Lambda_\Phi)$, 
to make the mass difference between  a CP-even scalar and a pseudoscalar bosons. 
Note that only $(H_1^\dagger H_2)^2 $ term can be multiplied by $U(1)_H$-charged 
operator $( \Phi / \Lambda_\Phi )^l $, whereas $(\Phi^\dagger \Phi/ \Lambda_\Phi^2)^n $ 
term can be multiplied to all terms because of $U(1)_H$ gauge invariance. 
If both are included, we would lose predictability because of too many nonrenormalizable
interactions.  Therefore we will choose $U(1)_H$ charge judiciously and $l=1$ is allowed 
by $U(1)_H$ gauge symmetry.  Then the leading nonrenormalizable operator would be 
dim-5 operator (the $c_{l=1}$ term),  and other operators with 
$(\Phi^\dagger \Phi/ \Lambda^2)^n $ would be at least dim-6 or higher. 
The  single unique dim-5 operator would lift up the mass degeneracy between a CP-even 
scalar and a pseudoscalar bosons by spontaneous $U(1)_H$ breaking.   
Therefore we do not lose predictability much even if we consider nonrenormalizable operators. 
We also give comment on how to realize $c_l$ in Sec. \ref{sec3}.
$(H_1^\dagger H_2) $ term may be also allowed, multiplying $( \Phi / \Lambda_\Phi )^l$
following the $U(1)_H$ charge assignment. This term may cause the nonzero VEV of $H_1$,
so that we define the charge assignment to forbid this term, as we see in Sec. \ref{sec3}.

After the EW and $U(1)_H$ symmetry breaking,
the VEVs of $H_i$ and $\Phi$ give the masses of the gauge bosons. 
In general,
the kinetic mixing between the SM gauge bosons and $\Hat{Z}_H$ may be 
generated as follows,
\begin{eqnarray}
{\cal L}_{\rm eff}  &=&  -\frac{1}{4} F^{\mu \nu}_Z F_{Z\mu \nu}  -\frac{1}{4} F^{\mu \nu}_{\gamma} F_{\gamma \mu \nu} -\frac{1}{4} F^{\mu \nu}_H F_{H\mu \nu} -\frac{\kappa_Z}{2} F_Z^{\mu \nu} F_{H\mu \nu}  -\frac{\kappa_{\gamma}}{2} F_{\gamma}^{\mu \nu} F_{H\mu \nu}   \nonumber \\
&& +\frac{1}{2} \Hat{M}^2_{Z} \Hat{Z}^{\mu} \Hat{Z}_{\mu} +\frac{1}{2} \Hat{M}^2_{Z_H} \Hat{Z}_H^{\mu} \Hat{Z}_{H\mu} +\Delta M^2_{Z_H Z} \Hat{Z}^{\mu} \Hat{Z}_{H\mu}. 
\end{eqnarray}
If $\langle H_1 \rangle$ is nonzero, $\Delta M^2_{Z_H Z}$ will be generated at the 
tree level.
When $\langle H_1 \rangle$ is zero, the tree-level mixing
of $Z$ and $Z_H$ does not exist, but the mixing may occur at the loop level,
as shown in the next subsection.

\subsection{Mass mixing and kinetic mixing at the one-loop level}
Before the EW and $U(1)_H$ gauge symmetry breaking, the local gauge symmetry forbids 
the mass mixing and the kinetic mixing among neutral gauge bosons, except for the 
kinetic mixing  $\kappa$ term in Eq.~(\ref{eq:lagrangian}).
If we assume that $\kappa$ is negligible, $\Hat{Z}_H$ does not couple with the SM fermions.

\begin{figure}[!t]
\begin{center}
{\epsfig{figure=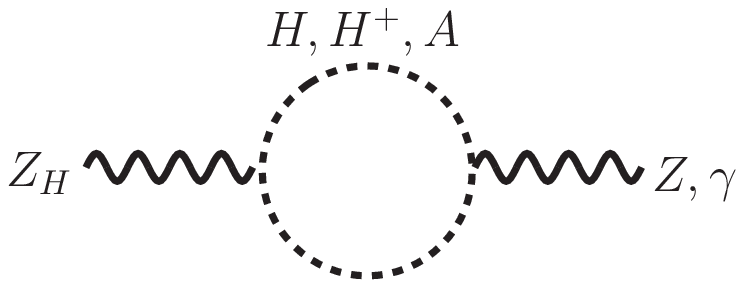,width=0.4\textwidth}}{\epsfig{figure=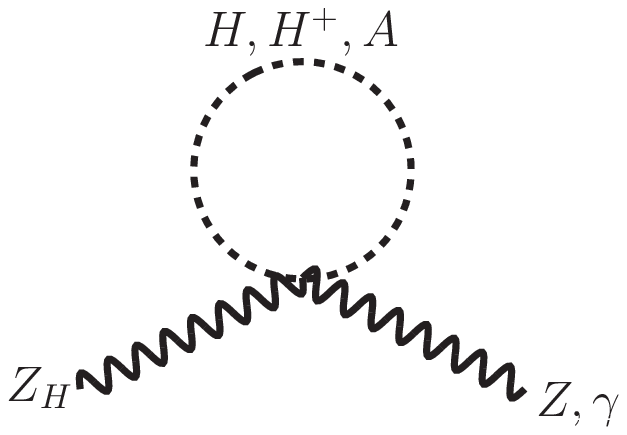,width=0.3\textwidth}}
\end{center}
\vspace{-0.5cm}
\caption{
The one-loop contribution of the extra scalars to $Z_H$-$Z$ and $Z_H$-$\gamma$ mixing in 
the IDMw$U(1)_H$ model.
}
\label{diagram1}
\end{figure} 
After the EW and $U(1)_H$ symmetry breaking, the mass mixing and the 
extra kinetic mixings, $\kappa_Z$ and $\kappa_{\gamma}$, will appear at the loop level, even if
$\kappa$ is negligible at the EW scale.
The one-loop contributions of the extra scalars 
to $\kappa_Z$ and $\kappa_{\gamma}$ are, for instance, given in Fig.~\ref{diagram1}. 
For nonzero $\langle H_1 \rangle$,  they are evaluated as
\begin{eqnarray}
\kappa_{Z}&=& \frac{q_{H}g_H e  c_W }{16\pi^2 s_W } \left \{ \cos^2 \beta Z_{WW} + \sin^2 \beta Z_{H^+H^+} -c^{A_l}_{\chi_1} c^{h_m}_{H_1} ( c^{A_l}_{\chi_1} c^{h_m}_{H_1}+ c^{A_l}_{\chi_2} c^{h_m}_{H_2}) Z_{A_lh_m} \right \}, \nonumber \\
&&  \\
\kappa_{\gamma}&=&  \frac{q_{H}g_H e  }{16\pi^2  } \left \{ \cos^2 \beta Z_{WW} + \sin^2 \beta Z_{H^+H^+} -c^{A_l}_{\chi_1} c^{h_m}_{H_1} ( c^{A_l}_{\chi_1} c^{h_m}_{H_1}+ c^{A_l}_{\chi_2} c^{h_m}_{H_2}) Z_{A_lh_m} \right  \}, \nonumber  \\
\end{eqnarray}
where $Z_{ab}$ is defined as
\beq
Z_{ab}=\frac{1}{3} \ln \left ( \frac{\Lambda^2}{m^2_{a}} \right) +\frac{1}{6} \frac{m_a^2-m^2_b}{m^2_a}.
\eeq
$\Lambda$ is the cut-off scale, but 
$\kappa_{Z,\gamma}$ are independent of $\Lambda$ because
the $\Lambda$ dependence is canceled in $\kappa_{Z,\gamma}$.
$c^{h_m}_{H_i}$ and $c^{A_m}_{\chi_i}$ are mixing angles
of the CP-even and CP-odd scalars, which are defined by
\begin{equation}
H^0_i=c^{h_m}_{H_i} h_m,
~\chi_i=c^{A_m}_{\chi_i} A_m,
\end{equation}
where $\{ h_m,A_l\}(\{H^0_i, \chi_i\})$ are the CP-even and CP-odd scalars in the mass (interaction) bases, respectively,  and the formulas for $\{ h_m,A_l\}(\{H^0_i, \chi_i\})$ are referred to Ref.~\cite{Ko-2HDM1}.
In the ordinary 2HDM, $\tan \beta$ is defined as $\tan \beta=\langle H^0_2 \rangle/\langle H^0_1 \rangle$, 
and $\langle H_1 \rangle=0$ corresponds to $\cos\beta = 0$.
In this limit, the physical neutral scalars can be expressed as
$h=h_2\cos\alpha - h_\Phi\sin\alpha$,
$H=h_1$,
and $\tilde{h}=h_2 \sin\alpha + h_\Phi\cos\alpha$, 
where $h_1$, $h_2$, and $h_\Phi$ are the CP-even neutral
components of $H_1$, $H_2$, and $\Phi$ after symmetry breaking, respectively,
and $\alpha$ is the mixing angle between $h_2$ and $h_\Phi$, where
$\Phi = (v_\Phi + h_\Phi + i \chi_\Phi)/\sqrt{2}$
and $H_2^0 = (v + h_2 + i \chi_2)/\sqrt{2}$.
$H_2^0 = (v + h_2 + i \chi_2)/\sqrt{2}$
The mass mixing, $\Delta M^2_{Z_H Z}$, is also induced at the one-loop level through the diagrams in Fig.~ \ref{diagram1}, 
\beq
\Delta M^2_{Z_H Z}=- \frac{q_{H}g_H  e }{16\pi^2s_Wc_W} F(m^2_{A_l},m^2_{h_m}) \{ c^{A_l}_{\chi_1} c^{h_m}_{H_1} ( c^{A_l}_{\chi_1} c^{h_m}_{H_1}+ c^{A_l}_{\chi_2} c^{h_m}_{H_2}) \}.  
\eeq

In the limit $\cos \beta \to 0$, which corresponds to $\langle H^0_1 \rangle \to 0$, the mixing parameters converge to
\begin{eqnarray} \label{eq:mixing1}
\kappa_{Z}&=& \frac{q_{H}g_H e  c_W }{16\pi^2 s_W } \left  \{ \frac{1}{3} \ln \left ( \frac{m^2_{A}}{m^2_{H^+}} \right ) -
\frac{1}{6} \frac{m^2_{A}-m^2_{H}}{m^2_{A}}\right \}, \\   \label{eq:mixing2}
\kappa_{\gamma}&=&  \frac{q_{H}g_H e  }{16\pi^2  }  \left  \{ \frac{1}{3} \ln \left ( \frac{m^2_{A}}{m^2_{H^+}} \right ) -
\frac{1}{6} \frac{m^2_{A}-m^2_{H}}{m^2_{A}}\right \} ,  \\   \label{eq:mixing3}
\Delta M^2_{Z_H Z}&=&- \frac{q_{H}g_H  e }{32\pi^2s_Wc_W} (m^2_{A}-m^2_{H}).  
\end{eqnarray}
Thus the mass and kinetic mixings of the gauge bosons are generated radiatively, 
even if the tree-level mass mixing is negligible.   

The $U(1)_H$ gauge coupling $g_H$ will be constrained by the collider experimental 
results through the couplings of $Z_H$ with the SM fermions  induced by the radiative 
corrections.  Even if $\kappa$ is zero at the cut-off scale ($\Lambda$),
it may become sizable at the low scale 
through the RG running.   In fact, the RG running correction can easily enhance 
the mixing in the case with large $g_H$.  For example, the RG flow of $\kappa_{\gamma}$ 
is estimated as
\beq
\kappa_{\gamma}( \mu ) \approx \kappa_{\gamma}(\Lambda)+ \frac{q_H}{48 \pi^2} g_H e \ln \left(\frac{\mu^2}{\Lambda^2} \right), 
\eeq
assuming that the running corrections of $e$ and $g_H$ are small.
Eventually, $g_H$ could be large if the masses of scalars are degenerate, but such large $g_H$  coupling would  enhance the kinetic mixing easily.
In our analyses, we set the cut-off scalar ($\Lambda$) at $1$ TeV, and include the radiative corrections to $\kappa$,   assuming $\kappa(\Lambda)=0$.

\subsection{Constraints on $Z_H$ coupling from the mixings}
The kinetic mixings and the mass mixing could be interpreted as the coupling of $Z_H$,
after changing the interaction basis to the mass basis.  Assuming that the mixing angles 
are small enough,  the couplings could be described as
\beq
  g_Z Z_{\mu} J_Z^{\mu} + e A_{\mu}  J_{\gamma}^{\mu} + \{ g_Z  (\xi -\kappa_Z) J_Z^{\mu}-  2e \kappa_\gamma J_{\gamma}^{\mu} \}Z_{H\mu},  
\eeq
where a mixing angle $\xi$ is given by the mass mixing,
\beq
\tan 2 \xi = \frac{2\Delta M^2_{Z_H Z}}{\Hat{M}^2_{Z_H} -\Hat{M}^2_Z}.
\eeq

The mixings among neutral gauge bosons are strongly constrained by the EWPOs and 
$Z'$ searches at high energy colliders, as discussed in Ref.~\cite{Ko-2HDM1}. 
If $Z_H$ is heavier than the center-of-mass energy of LEP ($209$ GeV),
we can derive the bound on the effective coupling of $Z_H$ 
\cite{Carena:2004xs, Alcaraz:2006mx}, depending on the $Z_H$ mass. 
The lower bound on $M_{Z_H}/g_H$ would be $\sim O(10)$ TeV \cite{Alcaraz:2006mx}. 
If $Z_H$ is lighter than $209$ GeV,
the upper bound of $Z_H$ coupling would be $O(10^{-2})$ in order that we avoid conflicts 
with the data from $e^+ e^- \to f^- f^+$ $(f=e, \mu)$ \cite{EWPO-PDG, Alcaraz:2006mx}.
If $Z_H$ is lighter than $M_Z$, the upper bound on the kinetic mixing is $\kappa \lesssim 0.03$ \cite{Chun:2010ve,Hook:2010tw}.  
In the very light $Z_H$ region ($100$ MeV $\lesssim$ $M_{Z_H} \lesssim 10$ GeV), the strong bound comes from the BaBar experiment,
$\kappa \lesssim O(10^{-3})$ \cite{Hook:2010tw}.

The upper bounds on the $Z_H$ production at the Tevatron and LHC 
are investigated in the processes, $p p(\overline{p}) \to Z_H X \to f \overline{f} X$ \cite{EWPO-PDG,Carena:2004xs,LHCZprime-ATLAS,LHCZprime-CMS}, and the stringent bound requires $O(10^{-3})$ times
smaller couplings than the $Z$-boson couplings around $M_{Z_H}=300$ GeV~\cite{LHCZprime-CMS}.

If we require the conservative bound $\sin \xi \lesssim 10^{-3}$, according to Ref.~\cite{LHCZprime-CMS},
we could estimate the upper bound in the large $\tan \beta$ case as follows, based on the Eqs. (\ref{eq:mixing1}),
(\ref{eq:mixing2}), and (\ref{eq:mixing3}),
\beq
 q_{H_1}g_H \left | \ln \left ( \frac{m^2_A}{m^2_{H^+}} \right ) \right | \lesssim  0.28,~ q_{H_1}g_H \left |   \frac{m^2_A-m^2_H}{\Hat{M}^2_{Z_H}-\Hat{M}^2_Z}  \right | \lesssim  0.43.
\label{gH}%
\eeq
Including the bounds from the scalar searches, we see the 
allowed region  for $g_H$ and $M_{Z_H}$ in Fig. \ref{fig1} in the case with  
$\langle H^0_1 \rangle=0$ $(\cos \beta=0)$.

\subsection{Constraints on the scalar bosons in the 2HDM with $\cos \beta=0$ }
In the limit $\cos \beta \to 0$, the new $U(1)_H$ gauge interaction could be large 
because the constraint from  the $\rho$ parameter is drastically relaxed.
At the tree level, the $\rho$ parameter constraint in general 
two Higgs doublet model with $U(1)_H$ symmetry (2HDMw$U(1)_H$) 
is given by \cite{Ko-2HDM}
\begin{equation}
 \{ q_{H_1} (\cos \beta )^2+ q_{H_2} (\sin \beta )^2 \}^2
 \frac{ g^2_H}{ g^2_Z}\frac{  M_{\Hat{Z}}^2 }{ M_{\hat{Z}_H}^2 - M_{\Hat{Z}}^2}
 \lesssim O(10^{-3}) ,
\label{eq:rho}%
\end{equation}
with $M_{\hat{Z}}^2 = g_Z^2 v^2$ and $M_{\hat{Z}_H}^2 = g_H^2 v^2 ( q_{H_1}^2 \cos^2\beta 
+ q_{H_2}^2 \sin^2 \beta ) + g_H^2 q_{\Phi}^2 v_\Phi^2$, where $q_{H_1,H_2,\Phi}$'s are the
$U(1)_H$ charges of $H_1, H_2$ and $\Phi$, respectively.

Now in the Type-I 2HDMs
the SM Higgs doublet and all the SM fields are $U(1)_H$ neutral
so that $q_{H_2} = 0$. 
In the IDM which we take into account for analysis in this paper,
$\langle H_1 \rangle = 0$, namely $\cos\beta = 0$. 
Then Eq.~(\ref{eq:rho}) implies that the $\Delta \rho$ constraint 
disappears at the tree level, and the $U(1)_H$ gauge 
coupling $g_H$ can be large.  
Also $Z_H$ gets its mass only from the VEV of 
the $\Phi$ field and can be relatively light. 
Furthermore, the scalar component of $H_1$ may be stable and could be a good cold 
dark matter candidate, as we discuss in the next section.
Below, we introduce the constraints on the extra scalars from the collider experiments 
and EWPOs.

\subsubsection{Constraints on the extra scalar bosons}
When $H_1$ does not develop 
a nonzero VEV, the scalar components in $H_1$, 
$(H,A,H^+)$,  are in the mass eigenstate and do not have the Yukawa couplings 
with the SM fermions.  This corresponds to the setup of the IDM~\cite{IDM,Ma}.
If $H$ is the lightest particle, $A$ and $H^+$ decay to $H$ and on-shell 
or off-shell $Z$, $Z_H$, $W^+$.  The constraints on $H$, $A$, and $H^+$ from the 
collider experiments have been widely discussed in the framework of the IDM~\cite{IDM-LEP}.
The search for multi leptons plus missing energy at LEP gives the lower bounds:
$m_{H^+} \gtrsim 90$ GeV and $m_{A} \gtrsim 100$ GeV with $m_A-m_H \gtrsim 8$ GeV.
The exotic $Z$ decay may be kinematically  forbidden by the condition, 
$m_A+m_H > M_Z$ in order not to change the decay width of the $Z$ boson.  

The mass differences between $m_{H^+}$ and $m_{A} (m_H)$
are also strongly constrained by the EWPOs,
and it has been studied in the Type-I 2HDMw$U(1)_H$ in the case 
with large scalar masses and generic $\tan \beta$~\cite{Ko-2HDM1}.
When $\tan \beta$ is large, the mass difference between the heavy CP-even 
scalar boson $(H)$ and the massive CP-odd scalar boson $(A)$
becomes small and the phase of $\Phi$ is eaten by $\Hat{Z}_H$.
$\tan \beta=0$ makes the masses degenerate in our 2HDM with $c_l=0$ 
in Eq. (\ref{eq:potential}), even after the EW and $U(1)_H$ symmetry breaking.
When $H$ and $A$ are degenerate and become CDM candidates, 
the $Z$ boson exchange diagram enhances the DM direct detection cross section,
for example, through $H + N \to A + N$, where $N$ is a nucleon,
and the CDM scenario would be excluded immediately by the XENON100 and LUX 
experiments.
A small mass difference $(\gtrsim O(100)$ keV) between $H$ and $A$ would be 
enough to suppress the direct detection cross section. 
In the 2HDMs with $Z_2$ symmetry, 
such a term is generated by $\lambda_5$ term, 
\begin{equation}
{\cal L}_{\lambda_{5}} = \frac{\lambda_5}{2} ( H_1^\dagger H_2 )^2 + H.c.
\end{equation}
which is clearly invariant under the usual discrete $Z_2$ symmetry,
$( H_1 , H_2 ) \rightarrow ( - H_1 , + H_2)$.  
However this terms is not allowed if we implement the discrete $Z_2$
symmetry to continuous $U(1)_H$ gauge symmetry at the renormalizable level. 
Still the effective $\lambda_5$ term may be induced by higher-dimensional 
operators integrating out heavy particles as we discuss below (see below, 
Sec.~\ref{sec:IDM}.).

\begin{figure}[!t]
\begin{center}
{\epsfig{figure=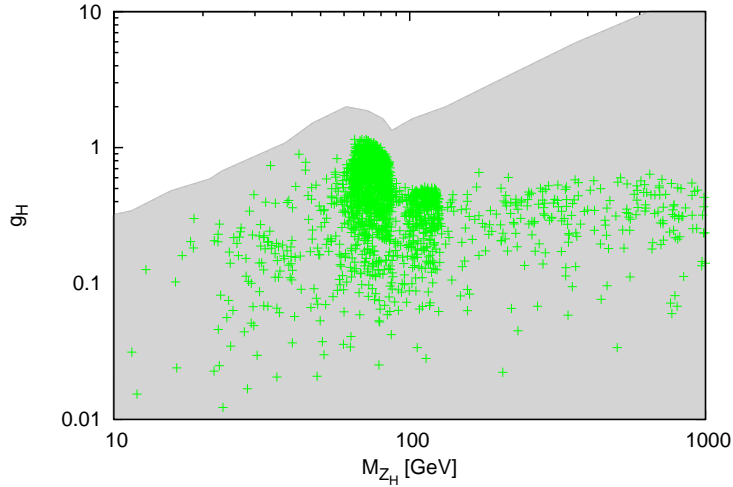,width=0.6\textwidth}}
\end{center}
\vspace{-0.5cm}
\caption{
Allowed region in the ($M_{Z_H} , g_H)$ plane with $\cos \beta=0$
in the IDMw$U(1)_H$.  
We have imposed the bounds from the collider search as well as EWPOs.  
The gray region satisfy the bounds from the thermal relic density for the CDM
at Planck
and the dark matter direct detection search at LUX.
The green points are also allowed by the dark matter indirect detection
search at Fermi-LAT.
}
\label{fig1}
\end{figure}

\subsubsection{Constraints from exotic SM-like Higgs decays}
The IDM condition $\cos \beta=0$ realizes the situation that $H_1$ completely 
decouples with the SM fermions,  so that the scalar components could be very light 
because of the relaxed experimental bounds, and the SM-like Higgs may decay to
the extra scalar bosons, as well as $Z_H$ and $\Phi$,
\[
h \to HH,A A, H^+H^-, \widetilde{h}\widetilde{h},ZZ_H,Z_HZ_H ,
\]
where $ \widetilde{h}$ is the CP-even scalar mainly from $\Phi$.
Including $H^+$ loop corrections, $h \to Z_H \gamma$ will open.
The extra scalars affect the signal strength $\mu$ of the SM-like Higgs boson 
at the LHC, where $\mu$ is defined in Ref.~\cite{Ko-2HDM1}. 
As discussed in Ref. \cite{Arhrib:2012ia},  $\mu_{\gamma \gamma}$ could be 
enhanced if $H^+$ is light. However, it is difficult to enhance $\mu_{ZZ}$ 
in the Type-I 2HDM.
When the extra fermions should be introduced to cancel the gauge anomaly,
which may appear according to the $U(1)_H$ charge assignments 
to the SM fermions, $\mu_{ZZ}$ could be enhanced from the contribution of 
color-charged  extra fermions to the $gg\to h$ production.
Now the ATLAS experiment shows a small enhancement for 
$\mu^{ggF+t\bar{t}h}_{\gamma\gamma}$ and 
$\mu^{ggF+t\bar{t}h}_{ZZ}$~\cite{atlashiggs}, 
but the CMS results for $\mu^{ggF+t\bar{t}h}_{\gamma\gamma}$ 
and $\mu^{ggF+t\bar{t}h}_{ZZ}$ are consistent with the SM prediction 
within $1\sigma$ error~\cite{cmshiggs}.
In this paper, we assume the SM-like Higgs boson and adopt the CMS results
for simplicity. 
Specifically, we impose the constraints on the signal strengths:
$\mu^{gg}_{\gamma \gamma}=0.70^{+ 0.33}_{- 0.29}$ 
and $\mu^{gg}_{Z Z}=0.86^{+ 0.32}_{- 0.26}$~\cite{cmshiggs,cmshiggs2}. 

When the $Z_H$ boson is light, 
the SM-like Higgs boson $h$ may decay into $Z_H Z_H$ or $Z Z_H$. 
If the mass of the dark matter $H$ is less than the half of the SM-like Higgs mass ($m_h$),
$h$ also can decay invisibly into $HH$. 
These exotic $h$ decays are strongly constrained by the search 
for the invisible and/or nonstandard Higgs decays at the LHC~\cite{invisible}.
We set the bound on the exotic Higgs decay to be~\cite{global,global_ko}
\begin{equation}
\frac{\sigma_\textrm{2HDM}^{Vh}}{\sigma_\textrm{SM}^{Vh}} \times
[\textrm{BR}(h\to Z Z_H )~\textrm{or}~
\textrm{BR}(h\to Z_H Z_H )] \le 0.69,
\end{equation}
where $\sigma_\textrm{SM,2HDM}^{Vh}$ is the cross section for 
the $Vh$ production in the SM and in the 2HDM, respectively,
and $\textrm{BR}(h\to Z_{(H)} Z_H )$ is the branching ratio
for the $h\to Z_{(H)} Z_H$ decay.

In Fig.~\ref{fig1}, the allowed region for $M_{Z_H}$ and $g_H$ is shown 
in gray color by taking into account the constraints from the $Z$-$Z_H$ mixing, 
EWPOs, search for exotic scalars, vacuum stability and unitarity, 
based on the above arguments and Ref.~\cite{Ko-2HDM1}.
For the numerical calculation we choose the following parameter spaces:
$20~\textrm{GeV} \le m_H, M_{Z_H} \le 1000$ GeV,
$90~\textrm{GeV} \le m_{H^+} \le 200$ GeV,
$63~\textrm{GeV} \le m_{\tilde{h}} \le 200$ GeV,
$0 \le \alpha \le 2\pi$,
$0 \le g_H \le 4\pi$,
$0 \le \lambda_1 \le 4\pi$,
$0 \le |\lambda_3|, |\tilde{\lambda}_1| \le 4\pi$,
and $-0.2 \le \lambda_5 \le 0$.
$\lambda_4$ is derived as 
$\lambda_4=2 (m_H^2-m_{H^+}^2)/v-\lambda_5$.

Since the SM fermions do not have the $U(1)_H$ charge, 
the relatively large $g_H$ is allowed as shown in Eq.~(\ref{gH}).
In the small $M_{Z_H}$ region, $g_H$ is $\sim O(0.1)$, but it can be $O(1)$ 
just below the $Z$ pole and much higher in the large $M_{Z_H}$ region.
We note that the gray region in Fig.~\ref{fig1} also satisfies
the constraints from the thermal relic density of the CDM, DM direct detection 
searches at the CDMSII, XENON100 and LUX experiments,
while the green points satisfy
the DM indirect detection search at Fermi-LAT as well as the relic density
and the DM direct detection search, 
which  we will discuss in the following section. 
In the small $M_{Z_H}$ region, the allowed values of $g_H$ do not change 
significantly even if we include the constraints from
the DM indirect detection search.
However, in the large $M_{Z_H}$ region, $g_H$ should be less than $O(1)$
when the constraints from the DM indirect detection search are taken into
account.
This is because one of the main channels for the annihilation of dark matters
in this region is $HH\to Z Z_H$, which would gives a stronger bound on $g_H$.

\section{ How to stabilize Dark matter in the IDMw$U(1)_H$}
\label{sec3}
In this section, we introduce a dark matter candidate in the IDMw$U(1)_H$ and 
compare predictions in the IDMw$U(1)_H$ with those in the IDMw$Z_2$~\cite{IDM}.
As we discussed in Sec.~\ref{sec2}, there is a $U(1)_H$ gauge boson 
as well as a $U(1)_H$-charged Higgs doublet in our
Type-I 2HDM with $U(1)_H$. 
If the Higgs doublet $H_1$ does not develop a VEV,
one of the scalar components of the doublet is a CDM candidate and 
the correct thermal relic density could be achieved.
First of all, let us discuss the stability of the dark matter in a generic $U(1)_H$ 
symmetric model in the Sec.~\ref{sec3-1}, and then discuss dark matter physics in the 
IDMw$U(1)_H$.

\subsection{General conditions for DM stability}
\label{sec3-1}
In general, we could build a gauge extension of the SM, such as $U(1)_H$.
As discussed in Ref.~\cite{Ko-2HDM}, not only additional Higgs doublets but also 
extra fermions may have to be introduced in order to 
satisfy the anomaly-free conditions, depending on the charge assignment.
The gauge symmetry would be spontaneously broken by extra scalars to avoid an extra 
massless gauge boson. 
Still, we could expect that there could be a residual local discrete 
symmetry after the spontaneous gauge symmetry breaking.
Then, if the extra fermions and/or scalars may respect the residual symmetry, 
their decays may be forbidden by the remaining local discrete  symmetry.

Let us discuss the generic $U(1)_H$ symmetric models with matters $\psi_I$ and $\Phi$
whose charges are $q_I$ and $q_{\Phi}$.  $\psi_I$ are $U(1)_H$-charged extra fields.
Simply, let us assume that $U(1)_H$ is broken only by a nonzero VEV of $\Phi$. 
Generally, the charges can be described as
\beq
\{q_{\Phi}, q_{I} \} = \left \{ \frac{n_{\Phi}}{N}, \frac{n_{I}}{N} \right \},
\eeq
where $n_I$, $n_{\Phi}$ and $N$ are integers, and all the charges
are irreducible fractions.
Now, we consider the case $\langle \Phi \rangle \neq 0$. 
In order to have a residual discrete symmetry after $U(1)_H$ breaking, the $\Phi$ 
field should be a singlet under the residual one after the $U(1)_H$ symmetry breaking. 
When we assume that $U(1)_H$ breaks down to $Z_{m}$ symmetry,
the integers $(m,m')$ can be defined by the relation,
\beq
 exp \left (i \frac{2 \pi m'}{m} \frac{n_{\Phi}}{N} \right)  \Phi =  \Phi.
\eeq
This could be satisfied by $(m,m')=(n_{\Phi},N)$ and 
we find that the residual symmetry is
\beq
U(1)_H \to Z_{|n_{\Phi}|}.
\eeq

In our 2HDM with $U(1)_H$, two fields $H_1$ and $\Phi$ are charged under $U(1)_H$.
If $H_1$ also develops a nonzero VEV, $\langle H_1 \rangle$ may also break the 
$Z_{|n_{\Phi}|}$. It depends on $q_{H_1}$, and the charge assignment may be fixed by
the $H^{\dagger}_1 H_2$ term,  namely by the operator 
\beq
\mu_n \Phi^n H^{\dagger}_1 H_2.
\eeq
This term should be forbidden because it will correspond to the tadpole term of 
$H_1$ which causes nonzero $\langle H_1 \rangle$ when $H_2$ and $\Phi$ break 
the EW and $U(1)_H$ symmetries by their nonzero VEVs.
Furthermore, if the $\mu_n$ term is allowed, the relation of the charges, $n_{H_1}=nn_{\Phi}$, is required, so that
$H_1$ is singlet under $Z_{|n_{\Phi}|}$,
as far as $n$ is an integer. 
In the next section, we discuss the IDMw$U(1)_H$
where only $\Phi$ breaks $U(1)_H$. 
$q_{H_1}$ is defined as $q_{H_1}/q_{\Phi}=n_{H_1}/n_{\Phi}$ is not an integer to realize a stable particle. 

\subsection{Toward the IDMw$U(1)_H$}
\label{sec:IDM}

One of the motivations for considering Type-I 2HDMs could be CDM, which
can be realized in the so-called inert 2HDM (IDMw$Z_2$),  
where one of the extra neutral scalar bosons could be a good CDM candidate.   

In the usual IDMw$Z_2$,  all the SM particles including the SM Higgs doublet with nonzero VEV are 
$Z_2$-even, whereas the other Higgs doublet without nonzero VEV is $Z_2$-odd \footnote{
Here $Z_2$ symmetry is presumed to be a global symmetry.}.
Then the scalar component of the $Z_2$-odd doublet could  
be stable and could be a good CDM candidate,  if the Higgs doublet does not get a nonzero VEV.     
One of the main issue in this type of CDM models is to generate mass differences among the charged, CP-odd, and CP-even components  of the $Z_2$-odd doublet,  in order to avoid strong constraints 
from the collider experiments and the direct dark matter searches. 
In the IDMw$Z_2$,  the $\lambda_4| H^{\dagger}_1 H_2|^2$  and $\lambda_5( H^{\dagger}_1 H_2)^2$ terms in the Higgs potential play an important role 
in the mass spectrum.  
Especially the $\lambda_5$ term 
shifts the pseudoscalar mass and thereby  suppresses kinematically  the 
interaction of the CP-even component with a nucleus ($N$) through the $Z$ exchange.   

In the IDMw$U(1)_H$, the discrete $Z_2$ symmetry is gauged to continuous local 
$U(1)_H$ symmetry,  so that  massless $U(1)_H$ gauge boson is predicted if the stability 
of dark matter is guaranteed by $U(1)_H$  Higgs gauge symmetry.
The $U(1)_H$ symmetry could be spontaneously broken, introducing SM singlet scalar 
$\Phi$ with a nonzero $U(1)_H$ charge $q_\Phi$. 
According to the general discussion in the previous subsection, the scalar components 
of $H_1$ could be stable if $\langle H_1 \rangle=0$ and $q_{H_1}/q_{\Phi}$ is  not an integer.  
However, the $U(1)_H$ symmetric Higgs potential generates an extra flat direction,
so that $H$ and $A$ tend to be degenerate. In fact, 
the $U(1)_H$ symmetry forbids the $\lambda_5$ term
at the tree level, which was the origin of the mass difference 
in the ordinary IDMw$Z_2$.  Without the $\lambda_5$ term, 
too large cross section for the direct detection of dark matter will be predicted, 
which is in serious conflict with the data.  In order to avoid this catastrophe,we generate the 
effective $\lambda_5$ term from  higher-dimensional operators, such as the $c_l$ term in Eq. (\ref{eq:potential}). 
Then the $\lambda_5$ term depends on $\langle \Phi \rangle$ in this scenario.

One simple way to realize higher-dimensional operators for an effective $\lambda_5$ 
term is to introduce an extra complex scalar $(\varphi)$ with a nonzero $U(1)_H$ charge and 
$\langle \phi \rangle = 0$.   Let us consider the following renormalizable potential among the scalar bosons: 
\begin{equation}
V_{\Phi}(|\Phi|^2,|\varphi|^2)+V_H(H_i,H^{\dagger}_i)+\lambda_{\varphi}( \Phi) \varphi^2+ \lambda_H (\varphi) H^{\dagger}_1H_2+h.c..
\end{equation}
where $\lambda_H$ and $\lambda_{\varphi}$ are  functions of $\varphi$ and $\Phi$, respectively, and  
respect the local $U(1)_H$ gauge symmetry as well as the SM gauge symmetries. 
Only $\Phi$ breaks $U(1)_H$ and only $H_2$ breaks the EW symmetry, due to nonzero values of 
$V_{\Phi}$ and $V_{H}$.  
Now we assume that $\varphi$ does not develop a nonzero VEV, and the direct coupling between 
$H_i$ and $\Phi$  is forbidden by a suitable choice of $q_{\Phi}$. 
If $U(1)_H$ is spontaneously broken  by the nonzero $\langle \Phi \rangle$, 
the squared mass difference $(\Delta m^2)$ between real and imaginary components 
of $\varphi$ will be generated by the $\lambda_{\varphi}$ term. Assuming that $\varphi$ is heavier than the 
EW scale and $\lambda_H=\lambda^0_H \varphi$,  $\lambda_5$ is induced effectively  at low energy 
when we integrate out the $\varphi$ field: 
\begin{equation}
\lambda_5\sim \frac{(\lambda^0_H)^2}{2}\frac{\Delta m^2}{m_{\varphi_R}^2m_{\varphi_I}^2},
\end{equation}
where $m_{\varphi_R}$ and $m_{\varphi_I}$ are the masses of the real and imaginary parts
of the complex scalar $\varphi$.  
After the $U(1)_H$ symmetry breaking,
the effective scalar  potential is the same as the one of the IDMw$Z_2$,
so that the small mass difference between $H$ and $A$ can be achieved, and
we can evade the strong bound from the direct detection search for dark matter.

\section{DM phenomenology in IDMw$U(1)_H$}
Base on the above argument and the setup, we discuss the dark matter physics in IDMw$U(1)_H$
and compare the results of the IDMw$U(1)_H$ with the ones of the IDMw$Z_2$.  
\subsection{Relic density and direct detection}
\label{sec3-3}
The most recent measurement for the DM relic density carried out
by the PLANCK Collaboration yields its value to be~\cite{PLANCK}
\begin{equation}
\Omega_\textrm{CDM} h^2 = 0.1199 \pm 0.0027.
\label{planck}%
\end{equation}
We impose that thermal relic densities calculated in the IDMw$Z_2$ and IDMw$U(1)_H$ 
satisfy this bound within 3$\sigma$ deviation,  assuming that the scalar dark matter 
in the IDMw$Z_2$ or IDMw$U(1)_H$ is the only CDM of the universe. 
In case we assume that there exist another dark matter particles
which contribute to the dark matter relic density,
we impose the bound $\Omega_\textrm{IDM} \le 0.1280$.

On the other hand,  dark matter may interact with atomic nuclei in a detector,
and it can be detected by underground experiments, such as the
DAMA~\cite{DAMA}, CoGeNT~\cite{CoGeNT}, CRESST~\cite{CRESST}, 
XENON~\cite{XENON10,XENON100}, CDMS~\cite{CDMS-II}, LUX~\cite{LUX} and etc.
The CoGeNT, DAMA, CRESST-II, and CDMS-II experiments show some excesses
in the light dark matter region with the dark matter mass of about 10 GeV
and the spin-independent cross section for the WIMP-nucleon scattering
is predicted to be $\sigma_{\rm SI} \sim 10^{-40}-10^{-42}$ cm$^2$~\cite{CDMS-II}.
It is interesting that the four experiments have some excesses
in the similar dark matter mass region, but the measured spin-independent cross section 
is different by about two orders of magnitude from each other. 
Also these signals are almost ruled out by the XENON100 and LUX experiments,
where the upper bound for the spin-independent cross section is
$\sim 10^{-45}$ cm$^2$ at $m_H \sim 33$ GeV and
and $\sim 10^{-44}$ cm$^2$ at $m_H \sim 10$ GeV~\cite{XENON10,XENON100,LUX}.
The positive signals might be accommodated with each other
while being reconciled with the XENON100 and LUX results
by introducing isospin-violating dark matter~\cite{isospin} 
or exothermic inelastic DM scattering.

\begin{figure}[!t]
\begin{center}
{\epsfig{figure=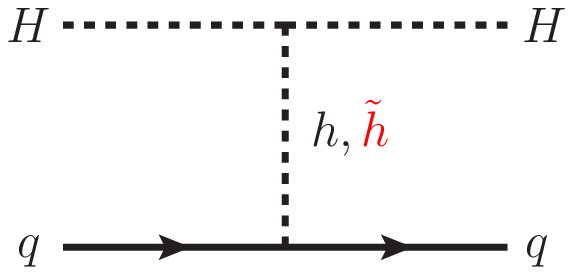,width=0.35\textwidth}} 
\hspace{0.2cm}
{\epsfig{figure=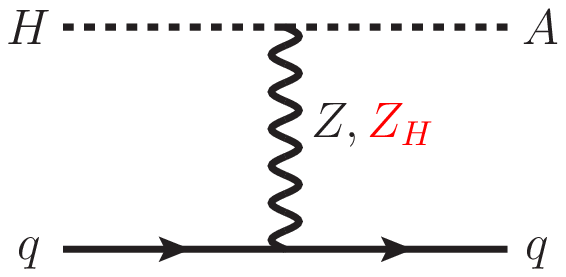,width=0.35\textwidth}}
\end{center}
\vspace{-0.5cm}
\caption{
Feynman diagrams for direct detection of DM in the usual IDMw$Z_2$ and additional ones in 
the gauged $U(1)_H$ model, IDMw$U(1)_H$.  Note that $\tilde{h}$ and $Z_H$ do not exist in the usual 
IDMw$Z_2$.
}
\label{diagram2}
\end{figure} 

In our model, we assume that the positive signals at low DM mass regions are excluded by the XENON100 
and LUX experiments and impose the bound for the spin-independent scattering cross section for the 
dark matter and nucleon from the LUX experiment.
In Fig.~\ref{diagram2}, we draw Feynman diagrams which dominantly
contribute to the direct detection of dark matter.
When the mass difference is negligible between $m_H$ and $m_A$,
the $Z$ and $Z_H$ exchange diagrams become dominant and
the spin-independent cross section for the dark matter candidate
and nucleon could exceed the bounds from the LUX experiment.
This problem is easily cured by the generation of the mass difference
between $m_H$ and $m_A$ with a sizable $\lambda_5$ term.
Then we can ignore $Z$ and $Z_H$ exchanges in Fig.~\ref{diagram2}, 
and the scalar ($h, \tilde{h}$) exchange is dominant 
in the direct cross section:  
\beq
\sigma_{SI}=\frac{\mu^2 f^2 m_N^2}{4 \pi m^2_H} \left \{ \left(\frac{\cos^2 \alpha}{m^2_h}+\frac{\sin^2 \alpha}{m^2_{\widetilde{h}}} \right ) (\lambda_3+\lambda_4+\lambda_5)+\left(\frac{1}{m^2_h}-\frac{1}{m^2_{\widetilde{h}}} \right ) \frac{v_{\Phi} \widetilde{\lambda}_1\cos \alpha \sin \alpha}{v} \right \}^2,
\eeq
where $f$ is the form factor, 
$\mu$ is the reduced mass of the DM-nucleon system, 
$m_N$ is the mass of the nucleon 
and $\alpha$ is the mixing angle 
between $h_\Phi$ and $H^0_2$.
We calculate the elastic cross section of the scattering of $H$ on atomic nuclei
by using micrOMEGAs, where the velocity of $H$ near the Earth, 
$v_H\approx 0.001c$~\cite{micromegas}.
In Fig.~\ref{fig2}, we show our predictions in this model. 
All the points in Fig.~\ref{fig2} pass the bound from the LUX experiment, 
as well as the collider experiments, which we described in Sec.~\ref{sec2}.

\begin{figure}[!t]
\begin{center}
{\epsfig{figure=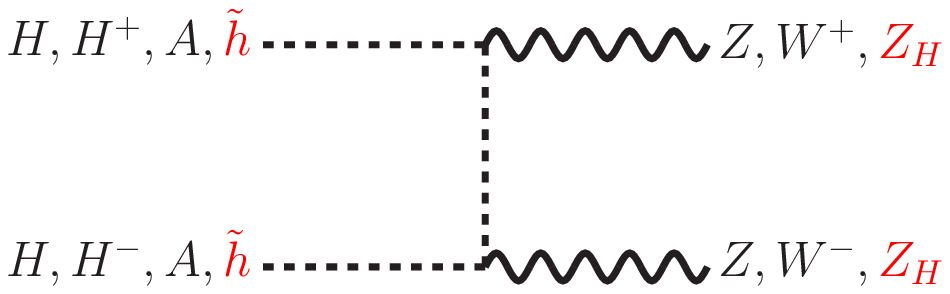,width=0.35\textwidth}} 
\hspace{0.2cm}
{\epsfig{figure=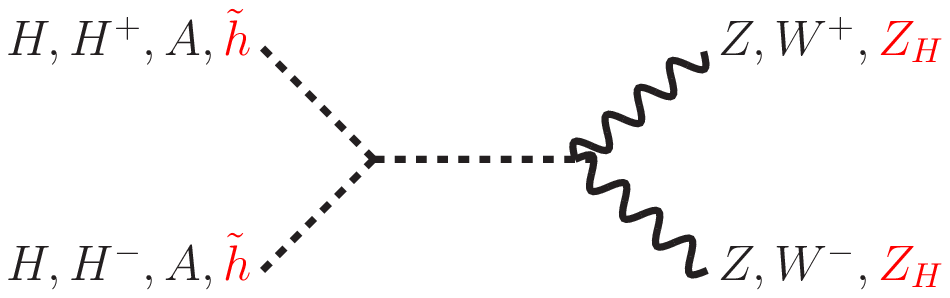,width=0.35\textwidth}}
\end{center}
\begin{center}
{\epsfig{figure=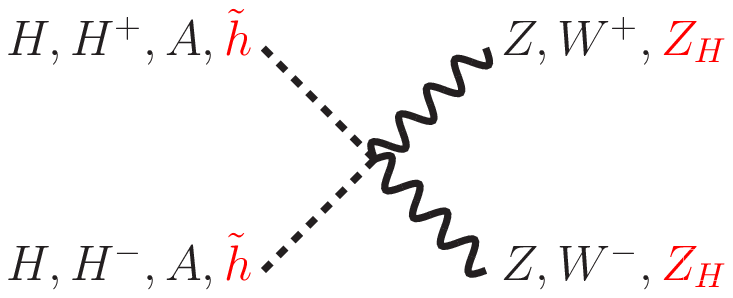,width=0.35\textwidth}}
\hspace{0.2cm}
{\epsfig{figure=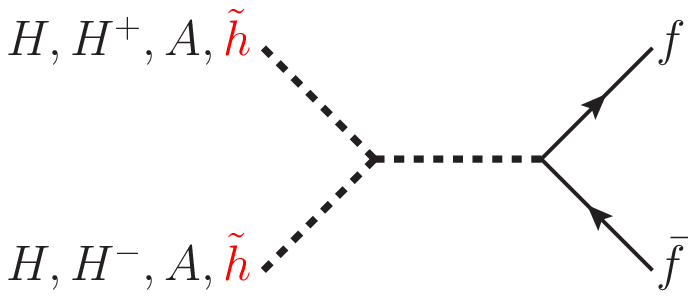,width=0.35\textwidth}}
\end{center}
\begin{center}
{\epsfig{figure=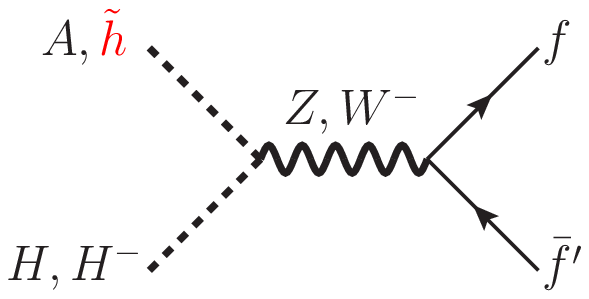,width=0.35\textwidth}}
\end{center}
\vspace{-0.5cm}
\caption{
Feynman diagrams for (Co)annihilation of dark matters in IDMw$Z_2$ and IDMw$U(1)$. The red ones are 
from $\Phi$ and $Z_H$ in the gauged $U(1)_H$ model, IDMw$U(1)$.  Dotted, solid and wavy lines denote 
spin-$0, 1/2$ and spin-1 particles, respectively.
}
\label{diagram3}
\end{figure}

\begin{figure}[!t]
\begin{center}
{\epsfig{figure=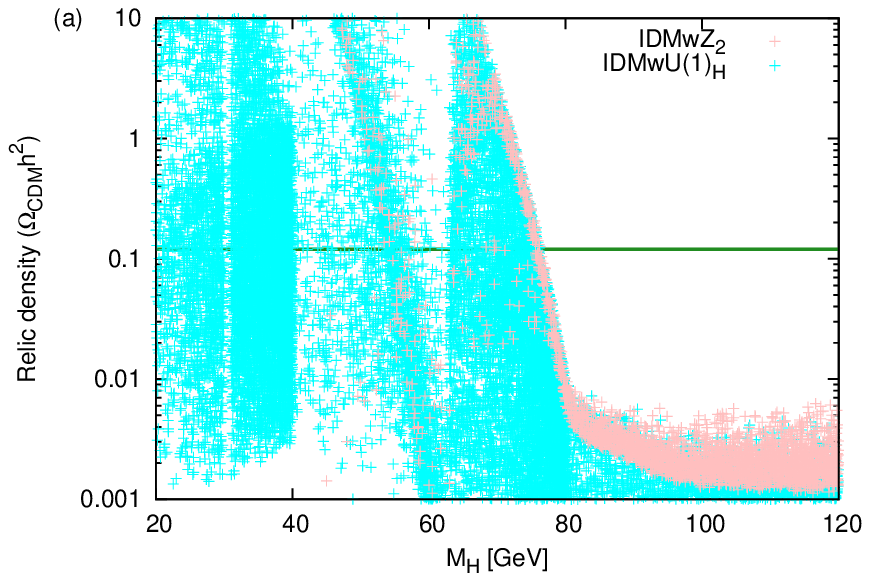,width=0.45\textwidth}}
{\epsfig{figure=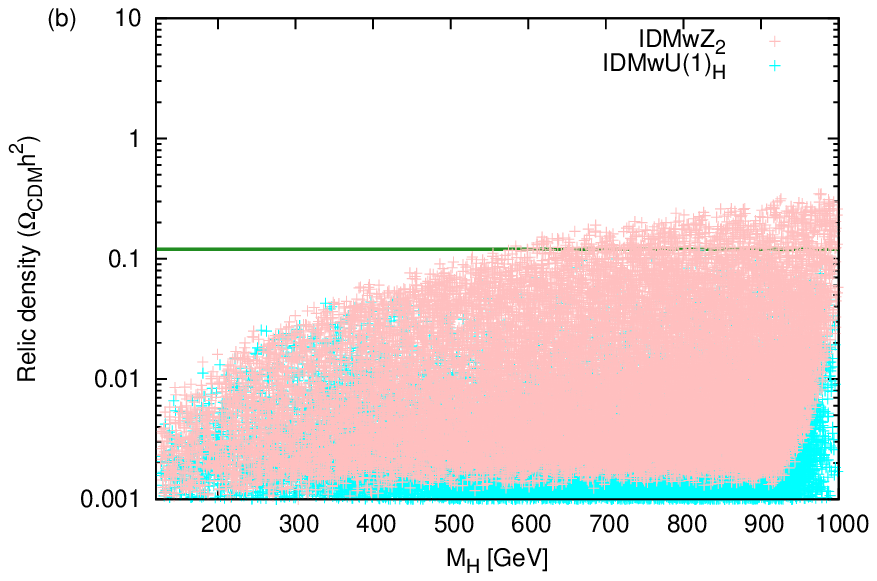,width=0.45\textwidth}}
\end{center}
\vspace{-0.5cm}
\caption{
$M_H$ and $\Omega h^2$ (a) in the light $H$ scenario and 
(b) in the heavy $H$ scenario.
}
\label{fig2}
\end{figure}

Thermal relic density, direct detections and indirect detection of DM 
in the IDMw$Z_2$ have been studied extensively in the literature (see, for example, \cite{IDM,Hambye:2009pw}).
In the IDMw$Z_2$, the extra scalars annihilate into two fermions through the 
SM-Higgs exchanging and two gauge bosons, as we see in Fig.\ref{diagram3}. 
There are two interesting scenarios: light dark matter ( $m_H \lesssim M_Z$) and
heavy dark matter ($m_H \gtrsim 500$GeV). 
In Fig.~\ref{fig2}, we show the relic densities (a) in the light $H$ scenario
and (b) in the heavy $H$ scenario, respectively.  The pink points correspond to 
the IDMw$Z_2$, whereas  the cyan points to the IDMw$U(1)_H$.
The horizontal line is the current value of the DM thermal relic density.

\subsubsection{The case of IDM with $Z_2$ symmetry}
Let us first explain the results of the IDMw$Z_2$.  
In the light dark matter scenario, the dark matter pair annihilation into the 
SM fermions ($HH\to f\bar{f}$) through the $s$-channel exchange of the SM Higgs 
boson $h$ makes a dominant contribution to the relic density in the range of 
$m_H \lesssim 50$ GeV.  However the direct detection experiments of DM~\cite{XENON10,LUX} 
and the invisible decay of the SM-like Higgs boson~\cite{global,global_ko} strongly 
constrain the $h-H-H$ coupling. Therefore the annihilation in that light mass region 
cannot be large enough to thermalize the scalar DM, and the light CDM scenario 
has been already excluded, as we see in Fig. \ref{fig2}.
 
In the $h$ resonance region, $m_H \sim m_h/2 \sim 60$ GeV,
the annihilation cross section is enhanced and the thermal relic density
could be below the current observation.
In the region with $m_H \gtrsim m_W$, the annihilation channels, 
$HH\to WW$ and $HH\to ZZ$ are open so that the IDMw$Z_2$ predicts
the small relic density in Fig.~\ref{fig2}.
The co-annihilation of $H$ and $A (H^+)$ also becomes relevant 
in the region with $40$ GeV $\lesssim m_H \lesssim 80$ GeV,
if the mass difference between $H$ and $A (H^+)$ is small.
In this case, 
the pair annihilation of $H^+ H^-$ or $AA$ could also be relevant
in the region $m_H \gtrsim 80$ GeV, and their contribution 
to the relic density could reach $60 \%$ of the relic density in a certain
parameter region.

In the heavy dark matter scenario, $HH \to ZZ,W^+W^-$ processes are 
very efficient and many channels like  $HH \to hh, t\bar{t}$ are open.
They can easily reduce the DM thermal relic density. 
As shown in Fig.~\ref{fig2} (b), the relic density is below the current
data in the region $m_H \lesssim 500$ GeV, and we have to consider extra DM 
species in order to account for the DM relic density in Eq.~(\ref{planck}). 
If we wish to accommodate
the current data  on the DM relic density within uncertainties in the IDMw$Z_2$, 
the DM mass $m_H$ should be greater than 500 GeV in the heavy dark matter
scenario. 
The pair annihilations of $H^+ H^-$ and $AA$ also contribute to the DM relic 
density in the region where the mass difference between $H^+, A$ and $H$
is small, but the ratio of their contribution to the relic density is less than $0.5$.
Typically the $H H\to W^+W^-$ process makes a dominant contribution  
to the DM relic density.  But in some parameter spaces 
the contribution of the pair annihilation of $H^+ H^-$ and $AA$ to the relic density
can be higher than that of the pair annihilation of $HH$. The process $HH\to hh$ can be 
relevant in some parameter spaces, but the ratio of the contribution is less than $0.3$.

\subsubsection{The case of IDM with local $U(1)_H$ symmetry}
Now let us discuss the result in the IDMw$U(1)_H$, the main theme of this paper.
In the IDMw$U(1)_H$, the new processes involving $Z_H$ could be dominant in the annihilation 
of the CDM.  For example, if $Z_H$ is lighter than $H$, the $HH\to Z_H Z_H$ process is open
and the $Z_H$ decays to the SM fermions through the $Z-Z_H$ mixing.  
We can see the relevant annihilation modes in Fig. \ref{diagram2}.
$\Phi$ may also work as the mediator if it is light. 
Fig.~\ref{fig2} (a) shows the relic density in the case $m_H \lesssim 126$ GeV. 
We also assume that $M_{Z_H}$ is also smaller than $126$ GeV
because the result in the limit $m_{Z_H} \gg m_H$ would not
be different from that in the IDMw$Z_2$.
If $M_Z + M_{Z_H} < 2 m_H$, then the $HH\to Z Z_H$ channel is also open and
it gives another contribution to the CDM relic density. The new channels involving 
$Z_H$('s) could be dominant  because of the weak bound on $g_H$ in Fig.~\ref{fig1}, 
and we could find the annihilations to $Z_H$ and SM gauge bosons play a role in 
decreasing the relic density  in Fig.~\ref{fig2} (a). 
Especially,  we can find many allowed points in the region with $m_H \lesssim 40 $ GeV 
in the IDMw$U(1)_H$ (cyan points in Fig.\ref{fig2} (a)), unlike the IDMw$Z_2$ case 
where the relic density becomes too large for  $m_H \lesssim 40 $ GeV. 
In this region, the gauge coupling $g_H$ of $U(1)_H$
prefers a small value, $g_H \lesssim 0.5$ as shown in Fig.~\ref{fig1}.

In the heavy $H$ scenario, the qualitative feature in the IDMw$U(1)_H$
is similar to that in the IDMw$Z_2$ as shown in Fig.~\ref{fig2} (b).
For $m_H\lesssim 500$ GeV, the predicted DM relic density is below the current
value and another dark matter is required to make up for the deficit. 
In the region $m_H \gtrsim 500$ GeV,
we can find the parameter regions which satisfy the current relic density.
In most parameter spaces, the predicted DM density in the IDMw$U(1)_H$
is slightly smaller than that in the IDMw$Z_2$.
This is because the new channels like $HH\to Z_H Z_H$  and $HH\to Z Z_H$ 
are open in the heavy CDM case, too,  so that they yield an extra contribution to 
the dark matter  density.

\begin{figure}[!t]
\begin{center}
{\epsfig{figure=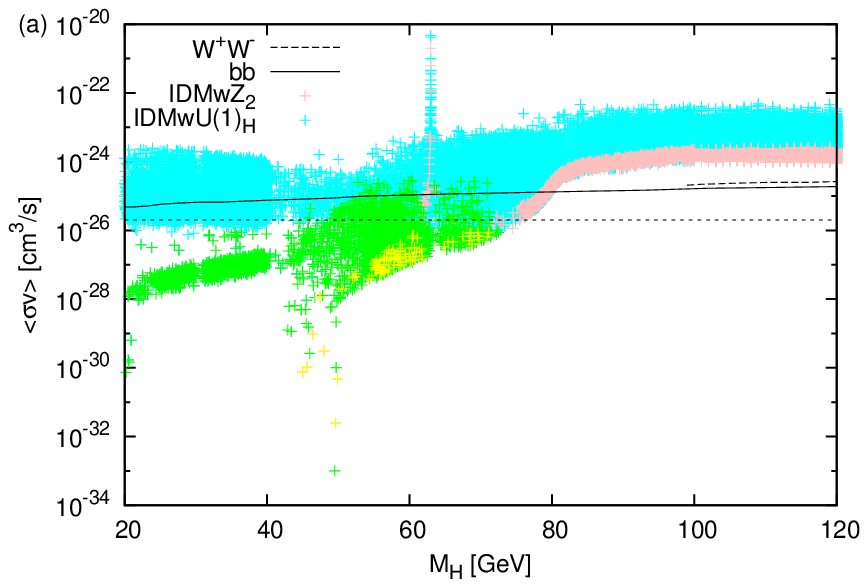,width=0.45\textwidth}}
{\epsfig{figure=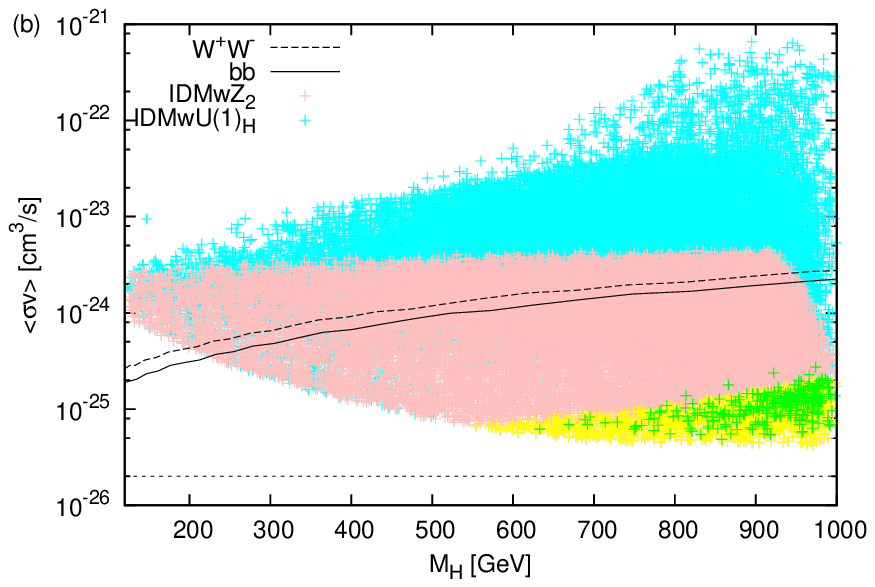,width=0.45\textwidth}}
\end{center}
\vspace{-0.5cm}
\caption{
$M_H$ in unit of GeV and $\langle \sigma v \rangle$ in unit of cm$^3/$s 
(a) in the light $H$ scenario and (b) in the heavy $H$ scenario. 
The pink and cyan points satisfy the constraints from the relic density and 
LUX experiments in the IDMw$Z_2$ and in the IDMw$U(1)_H$, respectively. 
The yellow and green points additionally satisfy the indirect constraints, 
$\Phi_\textrm{PP} \le 9.3\times 10^{-30} \textrm{cm}^3\textrm{s}^{-1} 
\textrm{GeV}^{-2}$ in the IDMw$Z_2$ and in the IDMw$U(1)_H$, respectively. 
}
\label{fig3}
\end{figure}

\subsection{Constraints from indirect detection}
\label{sec3-4}
In the IDMw$U(1)_H$, the light $Z_H$ gauge boson may open the new annihilation channels 
of the CDM: $HH \to Z_H Z_H, Z Z_H,\gamma Z_H$.  The $Z_H$ contributions could be dominant 
in the relic density because of the large $g_H$, and then astrophysical observations may give 
a crucial bound on the $Z_H$-dominant scenario,  as we discuss below.
In the heavy $H$ scenario, the $HH$ annihilation to the SM particles could be 
dominant because of gauge interactions.
In Fig.~\ref{diagram3}, we depict representative Feynman diagrams contributing
to the $HH$ annihilation.

In the galactic halo, the DM pair annihilation may occur and can be observed  
through the products of the annihilation : $\gamma$-rays, positrons, antiprotons, and etc.\footnote{If DM is not absolutely stable but lives much longer 
than the age of the universe, DM decays would produce similar observable effects in cosmic rays.
In this paper we consider the case that DM is absolutely stable due to the remnant discrete symmetry of
$U(1)_H$ gauge symmetry.}
This means that the detection of large signals for the $\gamma$-rays, positrons or antiprotons 
over astrophysical backgrounds would indicate indirect evidence for the dark matter annihilation.
There are some tantalizing excesses in the mono-energetic gamma ray signal around 
135 GeV observed by the Fermi-LAT~\cite{weniger,tempel,fermi-lat} 
and in the positron excesses at a few GeV $\sim$ TeV scale observed 
by PAMELA, Fermi, and AMS02 Collaborations~\cite{pamela,fermi09,ams02}.
However, it is still unclear that the origin of the excesses comes from
annihilation or decay of dark matter or astrophysical backgrounds~\cite{astrophy}.
In this paper, we assume that the excesses are originated in the astrophysical phenomena.
Then, indirect detection of dark matter plays a role
in constraining models.

Many experiments are devoted to measure the products of the dark matter
annihilation typically by observing cosmic rays
from dwarf spheroidal satellite galaxies~\cite{dwarf,dwarf2,dwarf3,FermiLAT25}
and galactic center~\cite{galatic}.
Especially, the dwarf spheroidal satellite galaxies
of the Milky Way are excellent targets for the detection of the dark matter
annihilation because of its large dark matter content and
low astrophysical backgrounds like a hot gas.
We calculate the velocity-averaged cross section for the dark matter
annihilation in the dwarf spheroidal galaxies by using micrOMEGAs~\cite{micromegas}.

The integrated $\gamma$-ray flux $\phi$, is divided two parts 
\begin{equation}
\phi_s (\Delta \Omega) = \Phi_\textrm{PP} \times J,
\end{equation}
where the second part is related to the dark matter distribution
in the galactic halo. 
Under the Navarro-Frenk-White density profile of dark matter~\cite{NFW}
the $J$ factor of each dwarf galaxy is given in Ref.~\cite{Jfactor}.
The quantity $\Phi_\textrm{PP}$ for the self-conjugate 
particles is defined by
\begin{equation}
\Phi_\textrm{PP} = \frac{\langle \sigma v\rangle}{8\pi m_H^2}
\int_{E_\textrm{min}}^{E_\textrm{max}} \sum_{f}^{} B_f 
\frac{d N_\gamma^f}{d E_\gamma} d E_\gamma,
\label{phipp}%
\end{equation}
where $\langle \sigma v \rangle$ is the velocity-averaged annihilation
cross section, $B_f$ is the branching ratio for each channel $XX\to f f^\prime$,
where $f$ and $f^\prime$ could be the SM particle or new particle,
and $d N_\gamma^f / d E_\gamma$ is the photon energy spectrum for
each annihilation channel. We note that the direct $\gamma$-ray production
from the dark matter annihilation which yields the mono-energetic signal
is loop-suppressed in both the IDMw$Z_2$ and the IDMw$U(1)_H$.
The photon energy spectrum depends on the decay patterns of intermediate states 
$f$ and $f^\prime$.

\subsubsection{The case of the IDM with $Z_2$ symmetry}
First, we consider the velocity-averaged annihilation cross section
in the IDMw$Z_2$ by using micrOMEGAs.
Figure~\ref{fig3} shows 
$\langle \sigma v \rangle$ in units of cm$^3/$s (a) in the light $H$ scenario
and (b) in the heavy $H$ scenario. 
The pink and yellow points correspond to the IDMw$Z_2$ while
the cyan and green points to the IDMw$U(1)_H$, respectively.
All the points satisfy thermal relic density and the direct detection 
constraint from the LUX experiments. And yellow and green points satisfy 
the additional constraints from indirect detections. 
The horizontal line around $3\times 10^{-26}$cm$^3/$s is the bound from the relic 
density when the $s$-wave annihilation of dark matters is dominant.
The upper region of this line would be allowed for the $s$-wave annihilation
dominant case.
The solid and dotted curves are the bounds from the Fermi-LAT experiment
by assuming that the dominant annihilation channel is 
$HH\to b\bar{b}$ and $HH\to W^+ W^-$.
The lower region of these curves is allowed if the dominant annihilation 
channel is $HH\to b\bar{b}$ or $HH\to W^+ W^-$.
We note that all the points in Fig.~\ref{fig3} satisfy
the constraints from the LUX experiment and their relic densities are below 
the current bound.

In the light $H$ scenario, the $HH\to b\bar{b}$ process can be dominant
in the Higgs-resonant region ($m_H\simeq 60$ GeV), whereas  
the $HH\to WW$ process becomes dominant both in the Higgs-resonant region
and in the region $m_H \simeq 80$ GeV. 
In these regions, the bounds from the Fermi-LAT experiment might constrain 
directly the IDMw$U(1)_H$ as well as the IDMw$Z_2$. From Fig.~\ref{fig3} (a),
we can find the allowed parameter spaces in the Higgs-resonant region, but
there is no allowed point at $m_H \gtrsim 80$ GeV.
We note that the region for $m_H \lesssim 40$ GeV is not allowed
in the IDMw$Z_2$ because of overclosing of the Universe.  
As we will see in the next subsection, this conclusion changes completely
when we extend the discrete $Z_2$ symmetry to local $U(1)_H$ symmetry because
new channels open, namely $HH \rightarrow Z_H Z_H , Z_H Z$ 
for on-shell or off-shell $Z$.

In the heavy $H$ scenario, the contribution of the $HH\to b\bar{b}$ process 
is negligible and the $HH\to W^+ W^-, ZZ$ processes become effective
in the IDMw$Z_2$. Depending on the coupling to the SM-like Higgs boson,
the $HH\to hh$ process could become dominant over the $HH\to W^+ W^-$ process.
In the $HH\to W^+ W^-$ dominant case, we can use
the bounds from the Fermi-LAT experiment.
We note that in the IDMw$Z_2$, $\langle \sigma v \rangle$ is less than about 
$3\times 10^{-24}$ cm$^3/$s. 
In the region $m_H \gtrsim 200$ GeV, there might be the parameter region 
allowed from the $\gamma$-ray observation by the Fermi-LAT when
the dominant DM annihilation channel is $HH\to W^+ W^-$.

\subsubsection{The case of IDM with local $U(1)_H$ symmetry}
In Fig.~\ref{fig3}, the cyan and green points depict 
the velocity-averaged annihilation cross section $\langle \sigma v \rangle$
versus the dark matter mass $m_H$ in the IDMw$U(1)_H$.   
We note that all points satisfy the bounds from the direct detection search
of dark matter at the LUX and the upper bound of the relic density, and 
the green points satisfy additional constraints from the indirect detection.

In the light $H$ scenario (Fig.~\ref{fig3}a), the allowed region 
in the IDMw$U(1)_H$ is much broader than that in the IDMw$Z_2$
because of additional channels like $HH\to Z_H Z_H$ and $HH\to Z Z_H$.
In the region $m_H \gtrsim 80$ GeV, the overall feature is the same as
in the IDMw$Z_2$ except that in the IDMw$U(1)_H$, $\langle \sigma v \rangle$
could be larger by an order of magnitude because of newly open channels 
such as $HH \rightarrow Z_H Z_H , Z_H Z$, etc.
In the Higgs-resonant region, $m_H \sim 60$ GeV, we find that some points
are allowed from the indirect detection search at the Fermi-LAT
if the dominant annihilation process is $HH\to b\bar{b}$.
As we already discussed in the previous subsection, 
there are no allowed points below $m_H \simeq 40$ GeV in the IDMw$Z_2$,
because the model predicts too large relic density.
However, in the $O(10)$GeV CDM scenario of the IDMw$U(1)_H$, 
we can find the allowed parameter spaces which
satisfy the constraints from the relic density and the direct detection search
of dark matter. 
In this region, the dominant annihilation channels are
$HH\to Z_H Z_H$ and $HH\to Z Z_H$, so that we cannot apply
the constraints from the indirect detection search of dark matter directly.

In the heavy $H$ scenario, the allowed parameter region from
direct detection search for dark matter and relic density 
in the IDMw$U(1)_H$ is broader than in the IDMw$Z_2$.
In this scenario, typically the $HH \to W^+ W^-, ZZ$ or
$HH \to hh$ processes could be dominant like the IDMw$Z_2$.
However, in the IDMw$U(1)_H$, 
the $HH\to Z_H Z_H$ or $HH\to Z Z_H$ could be dominant
processes over the $HH\to W^+ W^-, ZZ, hh$ processes, depending on the values
of the parameters in the model. It turns out that
the $HH\to Z_H Z_H$ process could become dominant at $m_H \gtrsim 500$ GeV,
whereas the $HH\to Z Z_H$ process could dominate over the other processes
at $m_H\gtrsim 200$ GeV.

\begin{table}[t]
\begin{center}
\begin{tabular}{|c||ccccc|cccccc|cc|}
\hline
&
~~$m_H$~~ & ~~$m_{H^+}$~~ & ~~$m_{\tilde{h}}$~~ & ~~$M_{Z_H}$~~ & ~~$m_A$~~& ~~$g_H$~~ 
 & ~~$\lambda_5$~~ & ~~$\lambda_1$~~ & ~~$\lambda_3$~~ & ~~$\tilde{\lambda}_1$~~ 
 & ~~$\alpha$~~
 & ~~$\Omega h^2$~~ 
 & ~~$\langle \sigma v \rangle_0$~~
\\ \hline
L1 &
$38.6$ &$189$ &$91.6$ &$39.2$ &$110$ &$0.33$ &$-0.174$ &$2.38$ &$1.09$ &$-2.37$ 
&$0.035$ &$0.113$ & $0.086$
\\ \hline
L2 &
$53.8$ &$194$ &$73.2$ &$29.6$ & $108$ &$0.215$ &$-0.144$ &$5.79$ &$1.09$ &$-1.59$ 
&$0.047$ &$0.117$ & $2.20$
\\ \hline
H1 &
$821$ &$822$ &$661$ &$985$ & $827$ &$0.235$ &$-0.164$ &$3.87$ &$0.15$ &$-0.429$ 
&$6.24$ &$0.119$ & $5.89$
\\ \hline
\end{tabular}
\caption{
\label{table1}%
{
Benchmark points in the IDMw$U(1)_H$. All the masses 
$m_H$, $m_{H^+}$, $m_{\tilde{h}}$, $M_{Z_H}$ and $m_A$ are in unit of GeV,  the mixing angle
$\alpha$ is in radian, and  $\langle \sigma v \rangle_0$ is in unit of $10^{-26}$ cm$^3/$s, respectively.
}
}
\end{center}
\end{table}

\subsubsection{$\Phi_\textrm{PP}$}
In the case that the $HH\to Z_H Z_H$ and $HH\to Z Z_H$
processes are dominant over other processes in the IDMw$U(1)_H$,
we cannot apply the bound from the indirect detection search of dark matter
at the Fermi-LAT, which assumes that annihilation into 
the $b\bar{b}$ or $W^+W^-$ pair is dominant.
The $\gamma$-ray spectrum strongly depends on the decay pattern
of particles produced from the dark matter annihilation.
The $Z_H$ boson dominantly decays into a pair of 
SM fermions through the $Z$-$Z_H$ mixing and its decay pattern is
similar to that of the SM $Z$ boson. In this case, 
the main source of the $\gamma$-ray
would be the $HH\to Z_H Z_H (Z) \to b\bar{b} b\bar{b}$ process.  
The $\gamma$-ray spectrum produced from this process would be different 
from the one from $HH\to b\bar{b}$. 
Therefore we should not apply
the bound from the Fermi-LAT shown in Fig.~\ref{fig3} directly.

In order to find out if our models can satisfy the constraints
from the indirect detection search at the Fermi-LAT,
we compute the quantity $\Phi_\textrm{PP}$ in Eq.~(\ref{phipp})
by using micrOMEGAs.
The range of the photon energy from 500 MeV to 500 GeV is summed.
For the comparison, we use the value in Ref.~\cite{dwarf2},
which was obtained by using the joint analysis of seven Milky Way dwarfs 
and Pass 7 data from the Fermi Gamma-ray Space Telescope.
A 95\% upper bound is $\Phi_\textrm{PP}=5.0^{+4.3}_{-4.5}\times 10^{-30}
\textrm{cm}^3\textrm{s}^{-1}\textrm{GeV}^{-2}$~\cite{dwarf2}.
In Fig.~\ref{fig3}, the green points satisfy this upper limit in the 
IDMw$U(1)_H$ while the yellow points satisfy this limit in the IDMw$Z_2$.
The pink and cyan points predict more $\gamma$-ray than the upper limit.

In the light $H$ scenario, only the Higgs-resonant region is allowed
in the IDMw$Z_2$, but the region $m_H \lesssim 40$ GeV is allowed
in the IDMw$U(1)_H$ too. In the Higgs-resonant region, 
$\langle \sigma v \rangle$ in some parameter spaces is below
the reference line for the $s$-wave annihilation dominant case 
at the decoupling temperature of dark matter.
There are two sources for the smaller $\langle \sigma v \rangle$
in the present universe. At the decoupling temperature, the velocity of
dark matter is $O(0.1)$, but the velocity at the halo is $O(0.001)$.
Another source is that the co-annihilation of $H H^+$ or $H A$
and pair annihilation of $AA$ or $H^+ H^-$ also contribute to the
relic density, but the co-annihilation does not appear in the halo at 
the current temperature.  In the region $m_H \lesssim 40$ GeV, only 
IDMw$U(1)_H$ can have a proper dark matter model to explain the bounds from 
the relic density, the direct and indirect detection searches of dark matter.
In this region, most of points predict smaller $\langle \sigma v \rangle$
than the reference line for the $s$-wave dominant annihilation case.
We find that the $HH\to Z_{(H)} Z_H$ process is dominant in this case
and $m_{Z_H}\approx m_H$. In this quasi-degenerate case,
there is a suppression factor in the phase space, which is proportional
to the velocity of dark matter. This can explain the gap of about 
$O(0.01\sim 0.001)$ in $\langle \sigma v \rangle$. 

In the heavy $H$ scenario, the allowed region appear at $m_H \gtrsim 500$ GeV
in both the IDMw$Z_2$ and IDMw$U(1)_H$. Most of points predict smaller
relic density than the observed one at the Planck experiment,
but some of them predict the exact relic density within $3\sigma$ uncertainties.
We note that the spin-averaged annihilation cross section in the halo 
at the zero temperature is over the reference line in the whole region.
We find that the contribution of $HH$ annihilation to the relic density
is at most $90 \%$ of the total relic density predicted in the model. 
Remaining the relic density is accounted for by the co-annihilation
of $H H^+$ or $H A$ and/or the pair annihilation of $AA$ or $H^+ H^-$,
which do not occur in the present halo.
This may explain the small gap between the reference line and 
the predicted $\langle \sigma v \rangle$ in the models.

\subsection{Three Benchmark Points for Illustration}

Since our model has many parameters,  we choose three benchmark points for which 
we can discuss the underlying physics in a more transparent manner.  
In Table~\ref{table1}, we show three benchmark points, all of which  
are safe for the astrophysical observations discussed above:  
the 1st one in the light dark matter region ($L1$),  the 2nd one in the $h$ resonance 
region ($L2$), and the 3rd one in the heavy dark matter region ($H1$).

At the $L1$ point, the dark matter mass is $m_H = 38.6$ GeV,
while the $U(1)_H$ gauge boson mass is $M_{Z_H}=39.2$ GeV, which is almost the 
same as $m_H$.  The coupling $\lambda_5 = -0.174$ generates the mass difference
between $A$ and $H$ with $m_A=110$ GeV, where the $HN\to AN$ process
through the $Z$ or $Z_H$ exchange is kinematically forbidden and we can evade the strong bound
from the direct detection search of dark matter. 
$\lambda_4=-0.957$ and
the spin-independent cross section for the WIMP-nucleon scattering is
$\sigma_\textrm{SI}=2.3\times 10^{-46}$ cm$^2$, which is below the LUX bound
at $m_H \sim 38$ GeV.  The corresponding relic density of DM is 
$\Omega h^2 = 0.113$, where $HH\to Z_H Z_H$ is the dominant channel for the correct thermal 
relic density.   The DM annihilation cross section  at the present universe is 
$\langle \sigma v \rangle_0 = 8.6\times 10^{-28}$ cm$^3/$s, where  the $HH\to Z Z_H$ process
contributes to the annihilation of $HH$ by 98\% and the $HH\to b\bar{b}$ process is about 2\%.

At the $L2$ point, $m_H=53.8$ GeV and $M_{Z_H}=29.6$ GeV.
$\lambda_5 = -0.144$, which leads to $m_A=108$ GeV.  
The mass difference between $H$ and $A$ is large enough to avoid the LUX bound
and the spin-independent cross section for the $H$ and nucleon is
$\sigma_\textrm{SI}=2.9\times 10^{-46}$ cm$^2$, which is below the LUX bound.
Here, $\lambda_4=-0.999$.
The relic density is  $\Omega h^2 = 0.117$, where the contribution of $HH \to Z_H Z_H$ 
to the relic density is 52\% and that of $HH\to b\bar{b}$ is 34\%.
Finally the DM annihilation cross section at the present universe is
$\langle \sigma v \rangle_0 = 2.20\times 10^{-26}$ cm$^3/$s.
The contribution of $HH\to Z Z_H$ to $\langle \sigma v \rangle_0 $ is 52\%
and that of $HH\to Z_H Z_H$ is 45\%.

At the $H1$ point, $m_H=821$ GeV and $M_{Z_H}=985$ GeV.
$\lambda_5 = -0.164$, which leads to $m_A=827$ GeV. 
In this case, $H^+$ and $A$ are almost degenerate to $H$, but still $m_A - m_H = 6$ GeV,    
which is large enough that one can evade the strong bound from the direct detection experiments.
$\lambda_4=0.086$ and
the spin-independent cross section for the $H$ and nucleon is
$\sigma_\textrm{SI}=6.1\times 10^{-46}$ cm$^2$, which is below the LUX bound.
The relic density is $\Omega h^2 = 0.119$ and a lot of processes contribute to the relic density 
because dark matter is much heavier than the SM particles ($m_H \gg m_W, m_Z, m_f$).
The contribution of each process of $HH\to Z Z_H, W^+ W^-$,
$H^+ H^- \to W^+ W^-, A Z_H$ is 11\%, 10\%, 9\%, and 9\%, respectively.
Besides these processes, many processes for $HH$, $H^+ H^+$, $H^+ H^-$, $H^+ H$
annihilation and so on contribute to the relic density.  The DM annihilation cross section 
at the present universe is $\langle \sigma v \rangle_0 = 5.89\times 10^{-26}$ cm$^3/$s.
The contributions of $HH\to Z Z_H$, $HH\to W^+ W^-$, and $HH\to Z Z$
to $\langle \sigma v \rangle$ are 34\%, 31\%, and 23\%, respectively.
Those of $HH\to hh, A W^+ W^-, t\bar{t}$ are less than 8\%.

In summary, three benchmark points have qualitatively different mass spectra and couplings. Still we find that
all these three points produce acceptable phenomenology for  the Higgs boson(s) and DM. 
The inert 2HDM with local $U(1)_H$ gauge symmetry has rich phenomenology 
due to new particles introduced by local $U(1)_H$ gauge symmetry. 
In particular, light DM ($\lesssim 60$ GeV) is still allowed  in the  gauged inert 2HDM (IDMw$U(1)_H$), 
unlike the usual inert 2HDM (IDMw$Z_2$) where DM below $\sim 60$ GeV is excluded, because of the strong 
bound from the direct detection cross section, thermal relic condition and the invisible decay of the SM Higgs. 
New particles and new interactions in the gauge $U(1)_H$ model help to cure these problems in the IDMw$Z_2$,  when suitable particle spectra and proper couplings are chosen in the IDMw$U(1)_H$. 

\section{Conclusion}
\label{sec4}
In this paper, we constructed inert 2HDM with local $U(1)_H$ gauge symmetry instead of the usual 
discrete $Z_2$ symmetry.  In this IDMw$U(1)_H$, the $U(1)_H$ gauge symmetry is spontaneously broken
by a nonzero VEV of a $U(1)_H$-charged SM-singlet scalar $\Phi$.  The $U(1)_H$-charged new Higgs
doublet $H_1$ neither couples to the SM fermions nor develops a nonzero VEV. The new Higgs doublet $H_1$ 
can be decomposed into a CP-even, CP-odd, and charged scalars ($H, A, H^\pm$, respectively), 
and one of the neutral scalar bosons (either $H$ or $A$) could be a good CDM  candidate because of 
the discrete $Z_2$ Higgs gauge symmetry, which is a remnant of the original $U(1)_H$ gauge symmetry 
spontaneously broken by nonzero $\langle \Phi \rangle$.  At the renormalizable level with these particle contents,  $H$ and $A$ are degenerate in mass, and the model is immediately excluded by the strong constraints from 
direct detection experiments on the $Z$ exchange to $H N \rightarrow A N$.   This problem can be solved
by lifting the mass degeneracy between the DM $H$ and the pseudoscalar $A$ by introducing another   $U(1)_H$-charged SM-singlet scalar $\varphi$ which does not develop nonzero VEV. 
This new singlet scalar $\varphi$ induces an effective $\lambda_5$ coupling as described in  Eq. (24), and would 
lift the degeneracy between $H$ and $A$. Then one can avoid the strong bound from the direct 
detection  cross section from the SM $Z$ exchange in $H N \rightarrow A N$. 

This scenario is a generalization of the well-known Inert Doublet Model, where the discrete symmetry $Z_2$ 
of IDMw$Z_2$ is replaced by local $U(1)_H$ gauge symmetry that is spontaneously broken into its $Z_2$ 
subgroup by nonzero VEV of $\Phi$.    
Like the usual IDMw$Z_2$, the new model we presented in this paper has many interesting features, 
such as the co-annihilation of the scalars and new channels for the DM pair  annihilations into $Z_H Z_H$, etc.

In the ordinary IDMw$Z_2$, there are two interesting CDM mass regions: 
$m_H \simeq 60$ GeV  and  $m_H \gtrsim 500$ GeV. In these regions, the correct DM relic 
density can be achieved without any conflict with the 
experimental results including the bound from the direct and indirect dark matter detections. 
Especially, the recent results from the indirect detection experiments may give strong bounds 
on the DM mass and its interactions.  We investigated the allowed region for the recent FERMI-LAT data 
in Sec. \ref{sec3-4}.   As we see in Fig.~\ref{fig3}, the FERMI-LAT data strongly constrains the DM annihilation 
cross section, so that the allowed regions are reduced in both of the light and heavy scenarios.
In fact, we found that $50-60$ GeV CDM scenario is only allowed in the light CDM case, 
because of the low annihilation cross section at the present universe temperature. 

In case of the IDMw$U(1)_H$, the nonzero $U(1)_H$ charge is assign to one Higgs doublet and 
one SM singlet,  and $U(1)_H$ is spontaneously broken by the singlet. 
The $U(1)_H$ breaks down to discrete symmetry, so that we could interpret the continuous 
symmetry as the origin of the $Z_2$ Higgs symmetry in the IDMw$Z_2$.
The additional massive gauge boson ($Z_H$) interacts with the extra Higgs doublet and 
weakly interacts with the SM particles through the mass and kinetic mixing with the SM gauge bosons in the IDMw$U(1)_H$.  If we assume the kinetic mixing between $U(1)_H$ and 
$U(1)_Y$ is negligibly small  before the EW and $U(1)_H$ symmetry breaking, 
we could expect that the $Z_H$ interaction could be sizable as we see in Fig. \ref{fig1}. 
In the dark matter physics, the $Z_H$ contribution might be dominant,  and it may make 
the IDMw$U(1)_H$ distinguishable from the IDMw$Z_2$.   In fact, we could find many allowed  
points for the bounds from the collider and dark matter experiments  below $m_H=60$GeV 
in the IDMw$U(1)_H$, where the IDMw$Z_2$ is totally excluded.
We also investigated the consistency with the recent FERMI-LAT data, and
we concluded that many points are still allowed in the light and heavy CDM mass regions.
In particular a new possibility opens up that the inert 2HDM can accommodate the 
$\gamma$-ray excess from the galactic center, if we promote the discrete $Z_2$ symmetry
to local $U(1)_H$ symmetry. This case will be discussed in detail elsewhere~\cite{GC}.

Our light CDM scenario predicts the exotic SM-Higgs decays: $h \to Z_H Z_H$ and $Z_H Z$.
$Z_H$ could decay to the SM fermions like the $Z$ boson, so that it may be possible
to observe $Z_H$ at LHC, as studied by the CMS collaboration \cite{CMS-Hig-13-010}. 
Developing the analysis of the SM-Higgs branching ratio,  
we may be able to draw the stronger bound on the exotic SM-Higgs decay according to 
the global fitting \cite{global,global_ko}.

\acknowledgments

We are grateful to Seungwon Baek, Wan-Il Park, and Yong Tang
for useful discussions and comments.
We thank Korea Institute for Advanced Study for providing computing resources 
(KIAS  Center for Advanced Computation Abacus System) for this work.
This work was supported in part by Basic Science Research Program through the
National Research Foundation of Korea (NRF) funded by the Ministry of Education Science
and Technology 2011-0022996 (CY), by NRF Research Grant 2012R1A2A1A01006053 
(PK and CY), and by SRC program of NRF funded by MEST (20120001176)
through Korea Neutrino Research Center at Seoul National University (PK).  
The work of YO is supported by Grant-in-Aid for Scientific research from the Ministry of Education, Science, Sports, and Culture (MEXT), Japan, No. 23104011.



\end{document}